\documentclass[11pt]{article}
\usepackage{amssymb,cite}
\makeatletter
\@addtoreset{equation}{section}
\makeatother

\def\ben{\begin{equation}}
\def\een{\end{equation}}

  \let\n=\nu

\let\C=\Chi

\def\nn{\nonumber} \def\bd{\begin{document}} \def\ed{\end{document}}
\def\ds{\documentstyle} \let\fr=\frac \let\bl=\bigl \let\br=\bigr
\let\Br=\Bigr \let\Bl=\Bigl
\let\bm=\bibitem
\let\na=\nabla
\let\pa=\partial \let\ov=\overline
\newcommand{\be}{\begin{equation}}
\newcommand{\ee}{\end{equation}}
\def\ba{\begin{array}}
\def\ea{\end{array}}
\def\ft#1#2{{\textstyle{{\scriptstyle #1}\over {\scriptstyle #2}}}}
\def\fft#1#2{{#1 \over #2}}
\def\del{\partial}
\def\vp{\varphi}
\def\sst#1{{\scriptscriptstyle #1}}
\def\oneone{\rlap 1\mkern4mu{\rm l}}
\def\td{\tilde}
\def\wtd{\widetilde}
\def\ie{\rm i.e.\ }
\def\dalemb#1#2{{\vbox{\hrule height .#2pt
        \hbox{\vrule width.#2pt height#1pt \kern#1pt
                \vrule width.#2pt}
        \hrule height.#2pt}}}
\def\square{\mathord{\dalemb{6.8}{7}\hbox{\hskip1pt}}}
\newcommand{\ho}[1]{$\, ^{#1}$}
\newcommand{\hoch}[1]{$\, ^{#1}$}
\newcommand{\bea}{\begin{eqnarray}}
\newcommand{\eea}{\end{eqnarray}}
\newcommand{\ra}{\rightarrow}
\newcommand{\lra}{\longrightarrow}
\newcommand{\Lra}{\Leftrightarrow}
\newcommand{\ap}{\alpha^\prime}
\newcommand{\bp}{\tilde \beta^\prime}
\newcommand{\tr}{{\rm tr} }
\newcommand{\Tr}{{\rm Tr} }
\def\0{{\sst{(0)}}}
\def\1{{\sst{(1)}}}
\def\2{{\sst{(2)}}}
\def\3{{\sst{(3)}}}
\def\4{{\sst{(4)}}}
\def\5{{\sst{(5)}}}
\def\6{{\sst{(6)}}}
\def\7{{\sst{(7)}}}
\def\8{{\sst{(8)}}}
\def\n{{\sst{(n)}}}
\def\cA{{{\cal A}}}
\def\cB{{{\cal B}}}
\def\cF{{{\cal F}}}
\def\cH{{{\cal H}}}
\def\tV{\widetilde V}
\def\tW{\widetilde W}
\def\tH{\widetilde H}
\def\tE{\widetilde E}
\def\tF{\widetilde F}
\def\tA{\widetilde A}
\def\im{{{\rm i}}}
\def\tY{{{\wtd Y}}}
\def\ep{{\epsilon}}
\def\vep{{\varepsilon}}
\def\R{\rlap{\rm I}\mkern3mu{\rm R}}
\def\bD{{{\bar D}}}
\def\R{\rlap{\rm I}\mkern3mu{\rm R}}
\def\bD{{{\bar D}}}
\def\R{{{\mathbb R}}}
\def\C{{{\mathbb C}}}
\def\H{{{\mathbb H}}}
\def\CP{{{\mathbb C}{\mathbb P}}}
\def\RP{{{\mathbb R}{\mathbb P}}}
\def\Z{{{\mathbb Z}}}
\def\bA{{{\mathbb A}}}
\def\bB{{{\mathbb B}}}
\def\bC{{{\mathbb C}}}
\def\bD{{{\mathbb D}}}
\def\bE{{{\mathbb E}}}
\def\bZ{{{\mathbb Z}}}
\def\Re{{{\mathfrak{Re}}}}
\def\Im{{{\mathfrak{Im}}}}
\def\cosec{{\,\hbox{cosec}\,}}
\def\Gm{{\Gamma_{\!\! -}}}
\def\Gp{{\Gamma_{\!\! +}}}
\def\stan{{standard }}
\def\nonstan{{supernumerary }}
\def\FF2{{ {}_{\sst 2}F_{\sst 1} }}
\def\FFF{{ {}_{\sst 3}F_{\sst 2} }}
\def\const{\rm constant}

\newcommand{\auth}{Muraari Vasudevan$^*$, Kory A. Stevens$^\dagger$ and Don N. Page$^{\dagger \dagger}$}

\begin{document}
\begin{flushright}

Alberta Thy 10-04\\

May\  2004
\end{flushright}

\vspace{10pt}

\begin{center}

{\large {\bf Separability of the Hamilton-Jacobi and Klein-Gordon Equations in
Kerr-de Sitter Metrics
            }}

\vspace{20pt}
\auth

\vspace{20pt}

\vspace{10pt}

{\it Theoretical Physics Institute,\\ University of Alberta,
Edmonton, Alberta  T6G 2J1, Canada}

{\it
  \medskip  {$^*$\rm E-mail: \texttt{mvasudev@phys.ualberta.ca}}
  }

{\it
  \medskip  {$^\dagger$\rm E-mail: \texttt{kstevens@phys.ualberta.ca}}
  }

{\it
  \medskip  {$^{\dagger \dagger}$\rm E-mail: \texttt{don@phys.ualberta.ca}}
  }


\vspace{40pt}

\underline{ABSTRACT}
\end{center}

We study separability of the Hamilton-Jacobi and massive Klein-Gordon equations
in the general Kerr-de Sitter spacetime in all dimensions. Complete separation of both equations is
carried out in $2n+1 $ spacetime dimensions with all $n$ rotation parameters equal, in which case the rotational
symmetry group is enlarged from $(U(1))^n$ to $U(n)$.  We explicitly construct the additional Killing vectors associated with the enlarged symmetry group which
permit separation. We also derive first-order equations of motion for particles
in these backgrounds and examine some of their properties.

\pagebreak
\setcounter{page}{1}

\tableofcontents
\addtocontents{toc}{\protect\setcounter{tocdepth}{2}}
\newpage

\section{Introduction}

Solutions of the vacuum Einstein equations describing black hole solutions in
higher dimensions are currently of great interest. This is mainly due to a
number of recent developments in high energy physics. Models of
spacetimes with large extra dimensions have been proposed to deal with several
questions arising in modern
particle phenomenology (e.g. the hierarchy problem). These models allow for the
existence of higher dimensional black holes which can be described classically.
Also of interest in these models is the possibility of mini black hole
production in high energy particle colliders which, if they
occur, provide a window into non-perturbative gravitational physics.

Superstring and M-Theory, which call for additional spacetime dimensions,
naturally incorporate black hole solutions in higher dimensions (10 or 11).
P-branes present in these theories can also support black holes, thereby
making black hole solutions in an intermediate number of dimensions physically
interesting as well. Black hole solutions in superstring theory are particularly
relevant since they can be described as solitonic objects. They provide important keys to understanding
strongly coupled non-perturbative phenomena which cannot be ignored at the Planck/string scale.

Astrophysically relevant black hole spacetimes are, to a very good
approximation, described by the Kerr metric \cite{Kerr}. One generalization of
the Kerr metric to higher dimensions is given by the Myers-Perry construction
\cite{MyersPerry}. With interest now in a nonzero cosmological constant, it is
worth studying spacetimes describing rotating black holes with a cosmological
constant. Another motivation for including a cosmological constant is driven by
the AdS/CFT correspondence. The study of black holes in an anti-de Sitter background could give rise to
interesting descriptions in terms of the conformal field theory on the boundary leading to
better understanding of the correspondence. The general Kerr-de Sitter metrics
describing rotating black holes in the
presence of a cosmological constant have been constructed explicitly in
\cite{GPLP}.

In this paper we study the separability of the Hamilton-Jacobi equation in these
spacetimes, which can be used to describe the motion
of classical massive and massless particles (including photons). We also
investigate the separability of the Klein-Gordon equation describing a spinless
field propagating in this background.
For both equations, separation is possible in some special cases due to the
enlargement of the dynamical symmetry group underlying these metrics. We
construct the separation of both equations explicitly in these cases. We also
construct Killing vectors, which exist due to the additional symmetry, and which
permit the separation of these equations. We also derive and study equations of motion for particles in these spacetimes.

\section{Construction and Overview of the Kerr-de Sitter Metrics}

A remarkable property of the Kerr metric is that it can be written in the
so-called Kerr-Schild \cite{KerrSchild} form, where the metric $g_{\mu \nu}$ is
given exactly by its linear
approximation around the flat metric $\eta _{\mu \nu} $ as follows:

\ben
ds^2 =g_{\mu \nu} dx^\mu dx^\nu  = \eta_{\mu \nu} dx^\mu dx^\nu
 + {2 M \over U} \, (k_\mu dx^\mu)^2\,,\label{ks1}
\een
where $k_\mu $ is null and geodesic with respect to both the full metric
 $g_{\mu \nu}$ and the flat metric $\eta _{\mu \nu}$.

The Kerr-de Sitter metrics in all dimensions are obtained in \cite{GPLP} by
using the de Sitter metric instead of the flat background $\eta _{\mu \nu}$,
with coordinates chosen appropriately to allow for the incorporation of the Kerr
metric via the
null geodesic vectors $k_{\mu}$. We quickly review the construction here.

We introduce $n=[D/2]$ coordinates $\mu_i$ subject to the constraint
\be
\sum_{i=1}^n \mu_i^2 =1\,,
\label{constraint}
\ee
together with $N=[(D-1)/2]$ azimuthal angular coordinates $\phi_i$,
the radial coordinate $r$, and the time coordinate $t$.  When the
total spacetime dimension $D$ is odd, $D=2n+1=2N+1$, there are $n$
azimuthal coordinates $\phi_i$, each with period $2\pi$.  If $D$ is
even, $D=2n=2N+2$, there are only $N=n-1$ azimuthal coordinates
$\phi_i$. Define $\epsilon$ to be 1 for even $D$, and 0 for odd $D$.

The Kerr-de Sitter metrics $ds^2$ in $D$ dimensions satisfy the Einstein
equation
\be
R_{\mu\nu} = (D-1)\, \lambda\, g_{\mu\nu}\,.
\ee
Define $W$ and $F$ as follows:
\be
W \equiv \sum_{i=1}^n \fft{\mu_i^2}{1+\lambda\, a_i^2}\,,\qquad
F\equiv \fft{r^2}{1-\lambda\, r^2}\, \,
  \sum_{i=1}^n \fft{\mu_i^2}{r^2+a_i^2}\,.\label{WFdef}
\ee
In $D$ dimensions, the Kerr-de Sitter metrics are given by
\be
ds^2 = d\bar s^2 + \fft{2M}{U}\, (k_\mu\, dx^\mu)^2\,,\label{kds}
\ee
where the de Sitter metric $d\bar s^2$, the null vector $k_\mu$, and the
function $U$ are now given by
\bea
d\bar s^2 &=& - W\,(1-\lambda \, r^2)\,
dt^2 + F\, dr^2 + \sum_{i=1}^n \fft{r^2 + a_i^2}{1+\lambda\, a_i^2}
\,\, d\mu_i^2 + \sum_{i=1}^{n-\epsilon} \fft{r^2 + a_i^2}{1+\lambda\, a_i^2}
\, \, \mu_i^2\, d\phi_i^2 \nn \\
&& \qquad +
 \fft{\lambda}{W\, (1-\lambda\, r^2)}\,
\Big( \sum_{i=1}^n \fft{(r^2 + a_i^2)\, \mu_i\, d\mu_i}{1+\lambda\, a_i^2}
   \Big)^2\,,\label{case}\\
k_\mu\, dx^\mu &=& W\, dt +  F\, dr -
\sum_{i=1}^{n-\epsilon} \fft{a_i\, \mu_i^2}{1+\lambda\, a_i^2} \,  d\phi_i \,,
\label{case2}\\
U &=& r^{\epsilon}\, \sum_{i=1}^n \fft{\mu_i^2}{r^2 + a_i^2}\,
\prod_{j=1}^{n-\epsilon} (r^2 + a_j^2)\,.\label{Udef}
\eea
In the even-dimensional case, where there is no azimuthal coordinate
$\phi_n$, there is also no associated rotation parameter; i.e., $a_n=0$.
Note that the null vector corresponding to the null one-form is
\be
k^\mu\, \del_\mu = -\fft{1}{1-\lambda\, r^2}\, \fft{\del}{\del t}
    + \fft{\del}{\del r} - \sum_{i=1}^{n-\epsilon}\, \fft{a_i}{r^2+ a_i^2}\,
\fft{\del}{\del \phi_i}\,.
\ee
This is easily obtained by using the background metric to raise and lower
indices rather than the full metric, since $k$ is null with respect to both
metrics.

For the purposes of analyzing the equations of motion and the Klein-Gordon
equation, it is very convenient to work with the metric expressed in Boyer-Lindquist
coordinates. In these coordinates there are no cross terms involving the
differential $dr$. In both even and odd dimensions, the Boyer-Lindquist form is obtained by means
of the following coordinate transformation:
\be
dt=d\tau + \fft{2M\, dr}{(1 - \lambda\, r^2)(V-2M)}\,,\qquad
d\phi_i = d\varphi_i - \lambda\,a_i\,d\tau +
\fft{2M\,a_i\, dr}{(r^2 + a_i^2)(V-2M)}\,.\label{coordtrans}
\ee
In Boyer-Lindquist coordinates in $D$ dimensions, the Kerr-de Sitter metrics
are given by

\bea
ds^2 &=& - W\, (1 - \lambda r^2)\,
d\tau^2 + \fft{U\, dr^2}{V-2M} +
\fft{2M}{U}\Bigl(d\tau - \sum_{i=1}^{n-\epsilon} \fft{a_i\, \mu_i^2\,
d\varphi_i}{
1 + \lambda\, a_i^2}\Bigr)^2 \nn\\
&&+ \sum_{i=1}^n \fft{r^2 + a_i^2}{1 + \lambda\, a_i^2}
\, d\mu_i^2 + \sum_{i=1}^{n-\epsilon} \fft{r^2 + a_i^2}{1 + \lambda\, a_i^2}\,
\mu_i^2\, (d\varphi_i-\lambda\, a_i\, d\tau)^2 \nn\\
&&+
\fft{\lambda}{W\, (1-\lambda r^2)}
\Big( \sum_{i=1}^n \fft{(r^2 + a_i^2)\mu_i\, d\mu_i}{
1 + \lambda\, a_i^2}\Big)^2 \,,\label{bl}
\eea
where $V$ is defined here by
\be
V\equiv  r^{\epsilon -2}(1-\lambda r^2)\prod_{i=1}^{n-\epsilon} (r^2 + a_i^2)
= \fft{U}{F}\,,\label{Vdef}
\ee
Note that obviously $a_n=0$ in the even dimensional case, as there is no rotation
associated with the last direction.

\section{Obtaining the Inverse Metric}

Note that the metric is block diagonal in the $(\mu _i)$ and the $(r,\tau
,\varphi _i)$ sectors and so can be inverted separately.

To deal with the $(r,\tau ,\varphi _i)$ sector, the most efficient method is to
use the Kerr-Schild construction of the metric. From (\ref{ks1}) and using the
fact that $k$ is null, we can write
\ben
g^{\mu \nu}=\eta ^{\mu \nu} -\frac{2M}{U}k^{\mu}k^{\nu} \,,
\een
where $\eta$ here is the de Sitter metric rather than the flat metric, and we
raise and lower indices with $\eta$. Since the null vector $k$ has no
components in the $\mu _i$ sector, we can regard the above equation as holding
true in the
$(r,\tau ,\varphi _i)$ sector with $k$ null here as well. Then we can explicitly
perform the coordinate transformation (\ref{coordtrans}) (or rather its inverse)
on the raised metric to obtain the components of $g^{\mu \nu}$ in
Boyer-Lindquist coordinates in the
$(r,\tau,\varphi _i)$ sector.

We get the following components for the $(r,\tau ,\varphi _i)$ sector of $g^{\mu
\nu}$:
\begin{eqnarray}
g^{\tau r}&=&g^{\varphi _i r}=0 \,, \nonumber \\
g^{rr}&=&\frac{V-2M}{U}\,, \nonumber \\
g^{\tau \tau}&=&Q-\frac{4M^2}{U(1-\lambda r^2)^2(V-2M)}\,, \nonumber \\
g^{\tau \varphi _i}&=&\lambda a_i Q -\frac{4M^2a_i(1+\lambda a_i
^2)}{U(1-\lambda r^2)^2 (V-2M)(r^2+a_i ^2)} -\frac{2M}{U}\frac{a_i}{(1-\lambda
r^2)(r^2+a_i ^2)}\,, \nonumber \\
g^{\varphi _i \varphi_j}&=& \frac{(1+\lambda a_i ^2)}{(r^2+a_i ^2)\mu _i ^2}
\delta ^{ij}+\lambda ^2 a_i a_j Q +Q^{ij}  \nonumber \\
&&+\frac{4M^2 a_i a_j (1+\lambda a_i^2) (1+\lambda a_j ^2)}{U(1-\lambda r^2)^2(V-2M)(r^2+a_i ^2)(r^2 +a_j ^2)}  \,, \label{rtfinv}
\end{eqnarray}
where $Q$ and $Q^{ij}$ are defined to be
\ben
Q=-\frac{1}{W(1-\lambda r^2)}-\frac{2M}{U}\frac{1}{(1-\lambda r^2)^2} \,,
\label{qlam}
\een

\begin{eqnarray}
Q^{ij}=\frac{-4M^2\lambda a_i a_j[(1+\lambda a_j ^2)(r^2+a_i ^2)+(1+\lambda
a_i^2)(r^2+a_j ^2)]}{U(1-\lambda r^2)^2 (V-2M)(r^2+a_i^2)(r^2+a_j
^2)}-\frac{2M}{U}\frac{a_ia_j}{(r^2+a_i ^2)(r^2+a_j ^2)}
\nonumber \\
-\frac{2M\lambda a_i a_j}{U(1-\lambda r ^2)} \left[ \frac{1}{(r^2 + a^2
_i)}+\frac{1}{(r^2 + a^2 _j )}\right] + \frac{4M^2 a_i a_j[(1+\lambda a_i
^2)+(1+\lambda a_j^2)]}{U(1-\lambda r^2)^2 (V-2M)(r^2+a_i^2)(r^2+a_j^2)} \,.
 \label{qij}
\end{eqnarray}

These results were compared to previously known ones in the case of $\lambda =0$
and showed agreement \cite{frolov1}. Also, we used the GRTensor
package for Maple explicitly to check that this is the correct inverse metric.

Note that the functions $W$ and $U$ both depend explicitly on the $\mu_i$'s. Unless
the $(r,\tau,\varphi_i)$ sector can be decoupled
from the $\mu$ sector, complete separation is unlikely. If however, all the
$a_i$'s are equal, then the functions $W$ and $U$
are no longer $\mu$ dependent (taking the constraint into account). With unequal
values of the rotation parameters $a_i$, separation does not seem to be
possible in this coordinate system, and it is likely that
a different coordinate system might be needed to analyze separability in those
cases. We will consider the case where all rotation parameters are equal: $a_i=a$.
Then we explicitly show separability. Note that since $a_n=0$ by definition for even
dimensional cases, we will restrict our attention to odd dimensional spaces. In
the discussions that follow, we explicitly set all rotation parameters equal, and assume that the spacetime
dimensionality is odd.

Note that the $\mu$ sector metric is completely diagonal upon assuming that the
rotation parameters are equal and upon imposing the constraint.
Consider the last term in equation (\ref{bl}) in the case of odd dimensions with all $a_i =a$. In this
case the term reads
\be
\fft{\lambda}{W\, (1-\lambda r^2)}\fft{(r^2 + a^2)}{(1 + \lambda a^2)} \Big(
\sum_{i=1}^n \mu_i d\mu_i\Big)^2 \,.
\ee
However, by differentiating the constraint (\ref{constraint}) we get $\sum _i \mu_i
d\mu_i=0$. Hence upon imposing the constraint this term vanishes from the
metric, and the corresponding term vanishes from the inverse metric (and thus in
the Hamilton-Jacobi equation.)

Now that the $\mu_i$'s are constrained by (\ref{constraint}), we can use
independent coordinates. Since the constraint describes
a unit $(n-1)$ sphere in $\mu$ space, the natural choice is to use spherical polar
coordinates. We write
\be
\mu_i= \left( \prod _{j=1}^{n-i} \sin\theta _j\right) \cos \theta _{n-i+1} \,,
\label{musphere}
\ee
with the understanding that the product is one when $i=n$ and that $\theta
_n=0$.
The $\mu$ sector metric can then be written as
\be
ds_{\mu}^2=\frac{r^2+a^2}{1+\lambda a^2} \sum _{i=1} ^{n-1} \left(\prod _{j=1}
^{i-1} \sin ^2 \theta _j\right)d\theta _i ^2 \,,
\ee
again with the understanding that the product is one when $i=1$.
This diagonal metric can be easily inverted to give
\be
g^{\theta _i \theta _j} = \frac{(1+\lambda a^2)}{(r^2+a^2)} \frac{1}{\left(\prod
_{k=1} ^{i-1} \sin ^2 \theta _k\right)}\delta _{ij} \label{muinv} \,.
\ee

\section{The Hamilton-Jacobi Equation and Separation}
The Hamilton-Jacobi equation in a curved background is given by
\be
-\frac{\partial S}{\partial l} = H = \frac{1}{2} g^{\mu \nu } \frac{\partial
S}{\partial x^{\mu}} \frac{\partial S}{\partial x^{\nu}} \,,
\ee
where $S$ is the action associated with the particle and $l$ is some affine
parameter along the worldline of the particle. Note that this treatment also
accommodates the case of massless particles, where the trajectory cannot be
parametrized by proper time.

Using (\ref{rtfinv}) and (\ref{muinv}), we write the Hamilton-Jacobi equation in
odd dimensions with all rotation parameters equal as

\[
	-2\frac{\partial S}{\partial l} = Q\left[ \frac{\partial S}{\partial \tau}
+\lambda a \sum _{i=1} ^{n} \frac{\partial S}{\partial \varphi _i}\right]^2 +
\frac{4M^2}{U(1-\lambda r^2)^2(V-2M)}\left[\frac{\partial S}{\partial \tau} -
\frac{a(1+\lambda a ^2)}{r^2+a ^2} \sum _{i=1} ^{n} \frac{\partial S}{\partial
\varphi _i}\right]^2
\]
\[
	-\frac{4M}{U(1-\lambda r ^2)}\frac{a}{(r^2+a ^2)} \sum _{i=1} ^{n} \frac{\partial
S}{\partial \tau} \frac{\partial S}{\partial \varphi _i}  + \frac{V-2M}{U}
\left(\frac{\partial S}{\partial r}\right)^2 -\frac{8M^2}{U(1-\lambda r
^2)^2(V-2M)}\left(\frac{\partial S}{\partial \tau}\right)^2
\]
\[
	+\sum _{ij=1} ^{n} Q^{ij} \frac{\partial S}{\partial \varphi _i}  \frac{\partial
S}{\partial \varphi _j} + \frac{(1+\lambda a ^2)}{(r^2+a^2)} \sum _{i=1} ^{n}
\frac{1}{\mu _i ^2} \left(\frac{\partial S}{\partial \varphi _i}\right)^2
\]
\begin{equation}
	+ \frac{(1+\lambda a ^2)}{(r^2+a^2)} \sum _{i=1} ^{n-1} \frac{1}{\left(\prod
_{k=1} ^{i-1} \sin ^2 \theta _k\right)}
	\left(\frac{\partial S}{\partial \theta_i}\right) ^2 \,.\label{HJ1}
\end{equation}
Note that here the $\mu_i$ are not coordinates, but simply notation defined by
(\ref{musphere}). The set of coordinates relevant to the problem
is $(\tau, r, \varphi _i, \theta _j)$. Note also that the functions $U,W,Q,$ and
$Q^{ij}$ are all now independent of the $\theta_i$; i.e., in the Hamilton-Jacobi
equation,
the $r$ sector has completely decoupled from the $\theta _i$ sector.

Now we can attempt a separation of coordinates as follows. Let
\begin{equation}
S=\frac{1}{2}m^2 l -E\tau + \sum_{i=1} ^n L_i \varphi _i +S_r (r)
+\sum _{i=1} ^{n-1} S_{\theta _i} (\theta _i)\,.
\end{equation}
$\tau$ and $\varphi _i$ are cyclic coordinates, so their conjugate momenta are
conserved. The conserved quantity associated with time translation is the energy
$E$, and those with rotation in the $\varphi _i$ are the corresponding angular
momenta $L_i$, all of which are conserved.  Applying this ansatz
to (\ref{HJ1}), we can separate out the overall $\theta$ dependence as

\be
J_1^2=\sum_{i=1}^{n} \left[ \frac{L_i ^2}{\left( \prod _{k=1}^{n-i} \sin
^2\theta _k\right) \cos ^2 \theta _{n-i+1}} \right] +
 \sum _{i=1} ^{n-1} \frac{1}{\left(\prod _{k=1} ^{i-1} \sin ^2 \theta
_k\right)} \left(\frac{dS_{\theta _i}}{d \theta _i}\right)^2 \,, \label{thetasep}
\ee
where $J_1^2$ is a constant. The separated $r$ equation is
\[
K=m^2(r^2+a^2)+Q(r^2+a^2)\left[-E+\lambda a \sum_{i=1} ^n L_i \right]^2 +
\frac{4MaE}{U(1-\lambda r^2)}\sum_{i=1} ^n L_i +
\]
\[
\frac{4M^2(r^2+a^2)}{U(1-\lambda r^2)^2(V-2M)}\left[E +
\frac{a(1+\lambda a ^2)}{r^2+a ^2} \sum _{i=1}^n L_i\right]^2
+\frac{(V-2M)(r^2+a^2)}{U}\left[\frac{dS_r}{dr}\right]^2
\]
\begin{equation}
-\frac{8M^2E^2(r^2+a^2)}{U(1-\lambda r^2)(V-2M)} + (r^2+a^2) \sum _{i,j=1} ^n Q^{ij}
L_i L_j \,, \label{rsep}
\end{equation}
where this separation constant is $K=-(1+\lambda a^2)J_1^2$. At this point the $(r,
\tau, \varphi _i)$ coordinates have been separated out. To show complete
separation of the
Hamilton-Jacobi equation we analyze the $\theta$ sector (\ref{thetasep}).

The pattern here is that of a Hamiltonian of a classical (non-relativistic)
particle on the
unit $(n-1)$ $\mu$-sphere, with some potential dependent on the squares of the
$\mu_i$. This can easily be additively separated following the usual
procedure, one angle at a time, and the pattern continues for all integers $n \ge 2$.

The separation has the following inductive form for $k=1,...,n-2$:
\begin{eqnarray}
&&J_k ^2 \sin ^2 \theta _k  -\frac{L^2 _{n-k+1} \sin ^2 {\theta _k}}{\cos ^2
{\theta _k}} -\sin ^2 {\theta _k} \left( \frac{d S_{\theta _k}}{d \theta
_k}\right)^2 = J^2 _{k+1} \,, \nonumber \\
&&J_{k+1} ^2 =\sum _{i=k+1} ^{n} \frac{L^2 _{n-i+1}}{\left(\prod
_{j=k+1}^{i-1}\sin ^2 \theta _j \right) \cos ^2 \theta _i } + \sum _{i=k+1}
^{n-1} \frac{1}{\left( \prod _ {j=k+1} ^{i-1} \sin ^2 \theta _j \right)} \left(
\frac{d S_{\theta _i}}{d\theta _i}\right) ^2 \label{thetasep1} \,.
\end{eqnarray}
The final step of separation gives
\be
J_{n-1} ^2 = \frac{L_2 ^2}{\cos ^2 \theta _{n-1}} + \frac{L_1 ^2}{\sin ^2 \theta
_{n-1}} + \left( \frac{d S_{\theta _{n-1}}}{d \theta _{n-1}}\right)^2 \,.
\label{thetasep2}
\ee

Thus, the Hamilton-Jacobi equation in odd dimensional Kerr-de Sitter space with
all rotation parameters $a_i=a$ has the general separation
\begin{equation}
S=\frac{1}{2}m^2 l -E\tau + \sum_{i=1} ^n  L_i \varphi _i +S_r (r) +\sum _{i=1} ^{n-1}
S_{\theta _i}(\theta _i) \,,
\end{equation}
where the $\theta _i$ are the spherical polar coordinates on the unit $(n-1)$
sphere. $S_r(r)$ can be obtained by quadratures from (\ref{rsep}), and the
$S_{\theta _i}$ again by quadratures from (\ref{thetasep1}) and
(\ref{thetasep2}).

\section{The Equations of Motion}
\subsection{Derivation of the Equations of Motion}

To derive the equations of motion, we will write the separated action $S$ from the Hamilton-Jacobi equation
in the following form:
\be
S=\frac{1}{2}m^2 l -E\tau + \sum_{i=1} ^n L_i \varphi _i +\int ^r \sqrt{R(r')} dr'
+\sum _{i=1} ^{n-1} \int ^{\theta _i} \sqrt{\Theta _i (\theta' _i)}d\theta ' _i
\ee
where
\be
\Theta_k = J_k ^2 -\frac{J^2 _{k+1}}{\sin ^2 \theta _k} -\frac{L^2
_{n-k+1}}{\cos ^2 \theta _k}\,,\qquad k=1,...,n-1\,,
\ee
\[
R=-J_1 ^2\frac{(1+\lambda a^2) U}{(V-2M)(r^2+a^2)} -\frac{QU}{(V-2M)}\left[-E
+\lambda a \sum _{i=1} ^n L_i \right]^2
\]
\[
- m^2\frac{U}{(V-2M)} -\frac{4M^2}{(1-\lambda r^2)^2 (V-2M)^2}\left[E
+\frac{a(1+\lambda a^2)}{r^2+a^2} \sum_{i=1}^n L_i\right]^2
\]
\be
- \frac{4MaE}{(V-2M)(r^2+a^2)} \sum_{i=1}^n L_i -\frac{8M^2E^2}{(1-\lambda
r^2)(V-2M)^2} - \frac{U}{(V-2M)}\sum_{i,j=1} ^n Q^{ij}L_i L_j \,,
\ee
where $Q$ and $Q^{ij}$ are functions of $r$ given in (\ref{qij}) (with all $a_i
=a$). For convenience, we define $J^2_n = L^2_1$. (Note that $J_n^2$ is obviously not a new conserved quantity. It is simply written this way
to facilitate the inductive definition given above for $\Theta _{n-1}$).

To obtain the equations of motion, we differentiate $S$ with respect to the
parameters $m^2,E,L_i, J^2_j$ and set these derivatives to equal
other constants of motion. However, we can set all these new constants of motion
to zero (following from freedom in choice of origin for the corresponding
coordinates, or alternatively by changing the constants of integration). Following this procedure, we get the
following equations of motion:
\begin{eqnarray}
\frac{dr}{dl}&=&\frac{(V-2M)\sqrt{R}}{U} \nonumber  \\
\frac{d\theta _i}{dl} &=& \frac{(1+\lambda a^2)\sqrt{\Theta _i}}{(r^2+a^2) (\prod _{j=1} ^{i-1}
\sin ^2 \theta _j)} \qquad i=1,...,n-1 \nonumber \\
\frac{d\tau}{dl}&=& 2Q(r^2+a^2)\left(E+\lambda a \sum _{i=1} ^n L_i\right) -\frac{4Ma}{U(1-\lambda
r^2)} \sum _{i=1} ^n L_i \nonumber \\
&&\hspace{-1cm} -\frac{8M^2(r^2+a^2)}{(1-\lambda r^2)^2 (V-2M)}\left(E+\frac{a(1+\lambda
a^2)}{(r^2+a^2)}\sum _{i=1}^n L_i\right)+ \frac{16 M^2 E (r^2+a^2)}{U(1-\lambda
r^2)(V-2M)} \,.\label{eqns}
\end{eqnarray}
We can obtain $n$ more equations of motion which give the $\frac{d\varphi
_i}{dl}$ in terms of the $r,\theta_j$ coordinates by differentiating $S$ with
respect to the angular momenta $L_i$. However,
these equations are not particularly illuminating, but can be written out
explicitly if necessary by following this procedure.

\subsection{Analysis of the Radial Equation}
The worldline of particles in the Kerr-de Sitter backgrounds considered above
are completely specified by the values of the conserved quantities $E,L_i, J^2
_j$, and by the initial values of the coordinates. We will consider
particle motion in the black hole exterior. Allowed regions of particle motion
necessarily need to have positive value for the quantity $R$, owing to equation
(\ref{eqns}). We determine some of the possibilities of the allowed motion.

At large $r$, the dominant contribution to $R$, in the case of $\lambda =0$, is
$E^2 -m^2$. Thus we can say that for $E^2<m^2$, we cannot have unbounded orbits,
whereas for $E^2>m^2$, such orbits are possible. For the case of nonzero
$\lambda$, the dominant term at large $r$ in $R$ (or rather the slowest decaying
term) is $\frac{m^2}{\lambda r^2}$. Thus in the case of the Kerr-anti-de Sitter
background, only bound orbits are possible, whereas in the Kerr-de Sitter
backgrounds, both unbounded and bound orbits may be possible.

In order to study the radial motion of particles in these metrics, it is useful
to cast the radial equation of motion into a different form. Decompose $R$ as a
quadratic in $E$ as follows:
\begin{equation}
R=\alpha E^2-2\beta E + \gamma \,,
\end{equation}
where
\begin{eqnarray}
\alpha &=& -\frac{QU}{V-2M} -\frac{4M^2}{(1-\lambda
r^2)^2(V-2M)^2}-\frac{8M^2}{(1-\lambda r^2)(V-2M)^2} \,, \nonumber \\
\beta &=& \left(\frac{QU\lambda a}{V-2M} + \frac{4M^2a(1+\lambda
a^2)}{(1-\lambda r^2)^2 (V-2m)^2(r^2+a^2)} +
\frac{2Ma}{(V-2M)(r^2+a^2)}\right)\sum ^n _{i=1} L_i \,, \nonumber \\
\gamma &=& -\frac{J_1 ^2(1+\lambda a^2) U}{(V-2M)(r^2+a^2)} -\frac{QU\lambda ^2
a^2}{V-2M} \left(\sum _{i=1}^n L_i\right)^2 -\frac{M^2 U}{V-2M} \nonumber \\
&& -\frac{4M^2a^2(1+\lambda a^2)^2}{(1-\lambda r^2)^2
(V-2M)^2(r^2+a^2)^2}\left(\sum _{i=1} ^n L_i\right)^2 -\frac{U}{V-2M} \sum _{ij=1}^n Q^{ij}
L_i L_j \,.
\end{eqnarray}

The turning points for trajectories in the radial motion (defined by the condition
$R=0$) are given by $E=V_{\pm}$ where
\be
V_{\pm} = \frac{\beta \pm \sqrt{\beta ^2 -\alpha \gamma }}{\alpha} \,.
\ee
These functions, called the effective potentials \cite{frolov1},
determine allowed regions of motion. In this form, the radial equation
is much more suitable for detailed numerical analysis for specific values of
parameters.

\subsection{Analysis of the Angular Equations}
Another class of interesting motions possible describes motion at a constant
value of $\theta _i$. These motions are described by the simultaneous equations
\begin{equation}
\Theta _i (\theta_i= \vartheta _i)=\frac{d\Theta _i}{d\theta _i}(\theta _i
=\vartheta _i) =0 \,,
\end{equation}
where $\vartheta _i$ is the constant value of $\theta _i$ along this trajectory.
These equations can be explicitly solved to give the relations
\begin{eqnarray}
\frac{J^2_{i+1}}{\sin ^4 \theta _i} &=& \frac{L^2 _{n-i-1}}{\cos ^4 \theta _i} \,,
\nonumber \\
J^2_i&=&\frac{J^2_{i+1}}{\sin ^2 \theta _i} +\frac{L^2_{n-i+1}}{\cos ^2 \theta
_i}\,,\qquad i=1,...,n-1\,,
\end{eqnarray}
where, as before, $J_n ^2=L_1 ^2$. Note that if $\vartheta_i=0$, then $J^2 _{i+1} =0$,
and if $\vartheta _i = \pi /2$, then $L_{n-i+1} ^2=0$.

Examining $\Theta _k $ in the general case, $\theta _k=0$ can only be reached if $J_{k+1}=0$, and
$\theta _k =\pi/2$ can be only be reached if $L_{n-k+1}=0$. The orbit will
completely be in the subspace
$\theta _ k=0$ only if $J_k ^2 = L^2 _{n-i+1}$ and will completely be in the
subspace $\theta _k =\pi/2$ only if $J_k ^2 = J_{k+1} ^2$.

Again these equations are in a form suitable for numerical analysis for specific
values of the black hole and particle parameters.

\section{Dynamical Symmetry}
The general class of metrics discussed here are stationary and ``axisymmetric";
i.e., $\partial / \partial \tau$ and $\partial / \partial \varphi _i$ are Killing
vectors and have associated conserved
quantities, $-E$ and $L_i$. In general if $\xi$ is a Killing vector, then $\xi
^{\mu} p_{\mu}$ is a conserved quantity, where $p$ is the momentum. Note that this quantity is first order in the momenta.

With the assumption of odd dimensions and equality of all the $a_i$'s, the
spacetime acquires additional dynamical symmetry and more Killing vectors are
generated. By setting the rotation parameters
$a_i$'s equal, we have complete symmetry between the various planes of rotation,
and we can ``rotate" one into another. The vectors that generate these
transformations are the required Killing vectors. We will construct these
explicitly.
Parametrize the rotation planes as follows:
\begin{eqnarray}
x_i=r\mu_i \cos \varphi _i = r\left( \prod _{j=1}^{n-i} \sin\theta _j\right) \cos \theta _{n-i+1}
\cos\varphi _i \,, \nonumber \\
y_i=r\mu_i \sin \varphi _i =r\left( \prod _{j=1}^{n-i} \sin\theta _j\right) \cos \theta _{n-i+1}
\sin\varphi _i\,,
\end{eqnarray}
again with the understanding that the product equals one when $i=n$ and that $\theta
_n =0$.

Define the rotation generators on the planes as
\begin{eqnarray}
L_{ab}=a\partial _{b}-b\partial _{a} \,,
\end{eqnarray}
where $a$ and $b$ can be any $x^i$ or $y^j$ . The case of $a=x^i,b=y^i$ for same $i$ is
not interesting, as it simply represents rotation in $\varphi_i$, which is already
known to generate a Killing vector. The $L_{ab}$ themselves are obviously not
Killing vectors (aside from the trivial cases just mentioned), but the
combinations
\be
\xi _{ij} = L_{x^i x^j} + L_{y^i y^j} \,, \qquad \rho _{ij} = L_{x^i y^j} + L_{x^j
y^i}\,,
\ee
are Killing vectors. Explicit expressions for these in polar coordinates in the
case of $n=2$ can be
found in \cite {frolov2} \cite{Kory}.

These additional Killing vectors exist, since the symmetry of the spacetime has
been greatly enhanced by the equality of the rotation
parameters. The $(U(1))^n$ spatial rotation symmetry symmetry, where each $U(1)$ is the rotational
symmetry in one of the planes, has been increased to a $U(n)$ symmetry. This
follows from the fact that we now have
the additional symmetry of being able to rotate planes into one another.

The separation constants $K$ in (\ref{rsep}) and $J_i^2$ in (\ref{thetasep}) are
conserved quantities that are quadratic in the associated momenta. So these
quantities must be derived from a
rank two Killing tensor $K^{\mu \nu}$\cite{Carter2}. We will work with the
$J_i^2$. (We can ignore $K$ since it only differs from
$J_1^2$ by a constant factor.) Any conserved quantity $A$ that is second order in
momenta is constructed from a
Killing tensor as
\be
A=K^{\mu \nu}p_{\mu}p_{\nu}=K^{\mu \nu}\frac{\partial S}{\partial
x^{\mu}}\frac{\partial S}{\partial x^{\nu}} \,. \label{killing1}
\ee

Since the Hamilton-Jacobi equation can be fully separated, we should be able to
construct Killing tensors explicitly. It turns out however that these Killing
tensors are not irreducible; i.e., they can be constructed as linear combinations
of tensor products of the Killing vectors
present due to the increased symmetry.

Comparing (\ref{thetasep}), (\ref{thetasep1}) and (\ref{thetasep2}) with
(\ref{killing1}), where the conserved quantities are $J_i^2$, we can obtain the
following Killing tensors:
\begin{eqnarray}
K_{n-1} ^{\mu \nu} &=& \frac{1}{\sin^2\theta _{n-1}} \delta
^{\mu}_{\varphi_1}\delta
^{\nu}_{\varphi_1} + \frac{1}{\cos^2\theta _{n-1}} \delta ^{\mu}_{\varphi_2}\delta
^{\nu}_{\varphi_2} + \delta ^{\mu}_{\theta_{n-1}}\delta ^{\nu}_{\theta_{n-1}}
\,, \nonumber \\
K_{k} ^{\mu \nu} &=& \frac{1}{\sin^2\theta _k} K_{k+1} ^{\mu \nu}
+\frac{1}{\cos^2\theta
_k} \delta ^{\mu}_{\varphi_{n-k+1}}\delta ^{\nu}_{\varphi_{n-k+1}} + \delta
^{\mu}_{\theta_k}\delta ^{\nu}_{\theta_k}\,, \qquad k=1,...,n-2\,,
\end{eqnarray}
which can be written as
\begin{eqnarray}
K_{n-k} &=&\sum_{i=1}^{k+1}\partial _{\varphi _i} \otimes \partial _{\varphi _i}
- \sum _ {i=1} ^ {k+1} \sum _{j=1} ^ {i-1} sym(\partial _{\varphi _i} \otimes
\partial _{\varphi
_j}) \nonumber \\ && +\sum_{i=1}^{k+1}\sum_{j=1}^{i-1}\xi_{ij}\otimes\xi_{ij}
+\sum_{i=1}^{k+1}\sum_{j=1}^{i-1}\rho_{ij}\otimes\rho_{ij}\,,\qquad k=1,...,n-1\,,
\end{eqnarray}
where $J_i ^2 = K_i ^{\mu \nu} p_{\mu} p_{\nu}$.

Therefore, as we can see from the form of the Killing tensors, they can explicitly
be obtained from quadratic combinations of the Killing vectors $\partial _{\varphi _i}$, $\xi_{ij}$,
and $\rho_{ij}$.

This is a demonstration of the fact that in this case
separation of the Hamilton-Jacobi equation is possible due to the enlargement
of the symmetry group in the case of all $a_i=a$.

\section{The Scalar Field Equation}
Consider a scalar field $\Psi$ with the action
\begin{equation}
S[\Psi]=-\frac{1}{2}\int d^Dx\sqrt{-g}((\nabla \Psi)^2+ \alpha R \Psi ^2 + m^2
\Psi ^2 ) \,,
\end{equation}
where we have included a curvature dependent coupling. However, in the
Kerr-(anti) de Sitter
background, $R = \lambda$ is constant. As a result we can trade off the
curvature coupling for a different mass term. So
it is sufficient to study the massive Klein-Gordon equation in this background.
We will simply set $\alpha=0$ in the following.
Variation of the action leads to the Klein-Gordon equation
\begin{equation}
\frac{1}{\sqrt{-g}}\partial _{\mu}(\sqrt{-g} g^{\mu \nu}\partial _{\nu} \Psi
)=m^2 \Psi \,.\label{KG1}
\end{equation}

As discussed by Carter \cite{Carter}, the assumption of
separability of the Klein-Gordon equation usually
implies separability of the Hamilton-Jacobi equation. Conversely, if the
Hamilton-Jacobi equation does not separate, the Klein-Gordon equation seems
unlikely to separate. We can also see this explicitly (as in the case of the Hamilton-Jacobi equation), since the $(r,\tau,\varphi_i)$ sector
has coefficients in the equations that explicitly depend on the $\mu_i$ except in the case of all $a_i=a$.
Thus, we will once again restrict our attention to the case of all $a_i=a$ in
odd dimensional spacetimes.

Once again, we impose the constraint (\ref{constraint}) and decompose the
$\mu_i$ in terms of spherical coordinates as in (\ref{musphere}). We calculate
the determinant of the metric to be
\begin{equation}
g=-\frac{r^2 (r^2+a^2) ^{2n-2}}{(1+\lambda a^2)^{2n}}
\prod_{j=1}^{n-1} \sin ^{4n-4j-2} \theta _j \cos ^2 \theta _j \,.
\end{equation}
For convenience we write $g=-PA$, where
\begin{equation}
P=\frac{r^2(r^2+a^2) ^{2n-2}}{(1+\lambda a^2) ^{2n}}\,,
\qquad A=\prod_{j=1}^{n-1} \sin ^{4n-4j-2} \theta _j \cos ^2 \theta _j \,.
\end{equation}
Then the Klein-Gordon equation in this background (\ref{KG1}) becomes
\[
	m^2 \Psi = Q\left[ \frac{\partial }{\partial \tau}
+\lambda a \sum _{i=1}^n \frac{\partial}{\partial \varphi _i}\right]^2 \Psi+
\frac{4M^2}{U(1-\lambda r^2)^2(V-2M)}\left[\frac{\partial }{\partial \tau} -
\frac{a(1+\lambda a ^2)}{r^2+a ^2} \sum _{i=1}^n \frac{\partial }{\partial
\varphi _i}\right]^2\Psi
\]
\[
	-\frac{4M}{U(1-\lambda r ^2)} \frac{a}{(r^2+a^2)} \sum _{i=1}^n  \frac{\partial ^2
\Psi}{\partial \tau \partial \varphi _i}-\frac{8M^2}{U(1-\lambda r
^2)^2(V-2M)}\left(\frac{\partial^2 \Psi}{\partial \tau ^2}\right)
\]
\[
	+\sum _{ij=1}^n Q^{ij}\frac{\partial^2 \Psi}{\partial \varphi _i \partial
\varphi_j}
	 + \frac{(1+\lambda a ^2)}{(r^2+a^2)}
	 \sum _{i=1}^n \frac{1}{\mu _i ^2}\left(\frac{\partial^2 \Psi}{\partial \varphi _i
^2}\right)
\]
\begin{equation}
	+\frac{1}{\sqrt{P}}\partial _r\left(\sqrt{P} \frac{(V-2M)}{U} \frac{\partial
\Psi}{\partial r}\right)+
	\frac{1}{\sqrt{A}}\sum_{i,j=1} ^ {n-1} \partial _{\theta_i}\left(\sqrt{A}
g^{\theta_i \theta_j}
\frac{\partial \Psi}{\partial \theta_j}\right) \,.
	\label{KG2}
\end{equation}
We attempt the usual multiplicative separation for $\Psi$ in the following form:
\begin{equation}
\Psi=e^{-iEt}e^{i\sum_i L_i \varphi _i} \Psi_{\theta}(\theta_1,...,\theta_{n-1})
\Phi_r (r) \,.
\end{equation}
Then the Klein-Gordon equation simplifies to give the following ordinary
differential equation in $r$ for $\Phi_r (r)$:
\[
	m^2 \Phi_r = -Q\left[ E-\lambda a \sum _{i=1}^n L_i\right]^2 \Phi_r
-\frac{4M^2}{U(1-\lambda r^2)^2(V-2M)}\left[E +
\frac{a(1+\lambda a ^2)}{r^2+a ^2} \sum_{i=1}^n L_i\right]^2\Phi_r
\]
\[
	-\frac{4MaE}{U(1-\lambda r ^2)(r^2+a^2)}\sum _{i=1}^n L_i
\Phi_r+\frac{8M^2E^2}{U(1-\lambda r
^2)^2(V-2M)}\Phi_r -\sum _{ij=1} ^n Q^{ij} L_i L_j \Phi_r
\]
\begin{equation}
	+\frac{1}{\sqrt{P}}\frac{d}{dr} \left(\sqrt{P} \frac{(V-2M)}{U}
\frac{d\Phi_r}{dr}\right)+
	\frac{(1+\lambda a^2)}{(r^2+a^2)} K_1 \Phi_r \,.
	\label{KG3}
\end{equation}
We have separated all the $\theta_i$ dependence into the separation constant $K_1$ given by
\be
K_1 = \frac{1}{\Psi_{\theta}}\sum_{i=1}^{n}\left[-\frac{L_i ^2}{\mu _i
^2}\right] + \sum_{i=1}^{n-1} \frac{1}{\Psi_{\theta}\sqrt{A} } \partial _{\theta
_i} \left( \sqrt{A} g^ {\theta _i \theta _i} \frac{\partial
\Psi_{\theta}}{\partial \theta _i}\right) \,,
\ee
where we have used the fact that $g^{\theta _i \theta _j}$ is diagonal, and that
the $\mu_i$ are functions
of the $\theta _j$ given by (\ref{musphere}).

Equation (\ref{KG3}) separates out the $r$ dependence of the Klein-Gordon
equation, and gives the function $\Phi_r(r)$ when the differential
equation is solved. We can also completely separate the $\theta_i$ sector.
Again, assume a multiplicative separation of the form
\begin{equation}
\Psi_{\theta}=\Phi_{\theta _1} (\theta _1)...\Phi _{\theta _{n-1}} (\theta _{n-1})\,.
\end{equation}
The $\theta$ separation then reads as
\begin{eqnarray}
K_1 &=& \sum _{i=1} ^{k-1} A_i + \frac{ K_k }{\prod _ {j=1} ^{k-1} \sin ^2\theta _j }\,, \quad k=1,...,n-1\,,
\end{eqnarray}
where
\begin{eqnarray}
A_i &=&  \frac{1}{\Phi _{\theta _i} \cos \theta _i \sin
^{2n-2i-1} \theta _i \prod _{k=1} ^{i-1} \sin ^2 \theta _ k}
\frac{d}{d\theta _i} \left( \cos \theta _i \sin ^{2n-2i-1} \theta _i \frac{d\Phi
_{\theta _i}}{d\theta _i} \right) \nonumber \\ && -\frac{L^2 _{n-i+1}}{\cos ^2 \theta _i \prod _{j=1} ^{i-1} \sin
^2{\theta _j}}  \,.
\end{eqnarray}

 Then we inductively have the complete separation of the $\theta _i$ dependence as
\begin{equation}
K_k  = \frac{K _{k+1}}{\sin ^2 \theta _k} -\frac{L^2 _{n-k+1}}{\cos ^2
\theta _k} + \frac{1}{\Phi _{\theta_k} \cos \theta _k \sin ^{2n-2k-1} \theta_k}
\frac{d}{d\theta _k} \left( \cos\theta _k \sin \theta _k \frac{d\Phi _{\theta
_k}}{d\theta _k} \right) \,,
\end{equation}
where $k=1,...,n-1$, and we use the convention $K_n = -L_1 ^2$.

As a result we can write the complete separation of the Klein-Gordon equation
(\ref{KG2}) in the Kerr-de Sitter background in odd dimensions with all rotation
parameters equal as
\be
\Psi=e^{-iEt}e^{i\sum_i L_i \varphi _i}
\Phi_{\theta_1}(\theta_1)...\Phi_{\theta_{n-1}}(\theta_{n-1}) \Phi_r (r) \,,
\ee
where $\Phi(r)$ is obtained from (\ref{KG3}), and the $\Phi_{\theta _ i}$'s are
the decomposition of the $\mu$ sector into eigenmodes in independent coordinates
$\theta _i$ on the $\mu$ sphere.

Note that the separation of the Klein-Gordon equation in this geometry is again
due to the fact that the symmetry of the space has been enlarged. (We can
explicitly see the role of the Killing
vectors again in the separation of the $r$ equation from the $\theta$ sector in a
very similar fashion to that in the Hamilton-Jacobi equation \cite{Carter}).

\section*{Conclusions}
We studied the separability properties
of the Hamilton-Jacobi and the Klein-Gordon equations in the Kerr-de Sitter backgrounds.
Separation in Boyer-Lindquist coordinates seems to be possible only for the case
of an odd number of spacetime dimensions with all rotation parameters equal. This is possible due to the
enlarged dynamical symmetry of the spacetime. We derive expressions for the
Killing vectors that correspond to the additional symmetries. We also show that integrals of motion are
obtained from reducible Killing tensors, which are themselves constructed
from the angular Killing vectors. Thus we demonstrate the separability of the
Hamilton-Jacobi and the Klein-Gordon equations as a direct consequence of the enhancement of symmetry.
We also derive first-order equations of motion for classical particles in these backgrounds,
 and analyze the properties of some special trajectories.

Future work in this direction could include finding a suitable coordinate system
to permit possible separation in an even number of spacetime dimensions. Different coordinates might also be required to study the cases of unequal
rotation parameters, since separation does not seem likely in Boyer-Lindquist coordinates.
The study of higher-spin field equations in these backgrounds could also prove to be of great interest,
particularly in the context of string theory. Explicit numerical study of the equations of motion for specific values of the black hole parameters
could lead to interesting results.

\section*{Acknowledgments}

We are grateful to Gary Gibbons for providing a copy of earlier work by Rebecca Palmer\cite{Palmer} that made progress toward separation in higher
dimensional Myers-Perry metrics. Our research was supported in part by the Natural Science and Engineering Research Council of
Canada.

\end{document}